%% file: npfp.tex
\documentclass{article}
\usepackage{epsfig}
\usepackage{axodraw}

\newcommand{\sheptitle}{Neutrino physics and the flavour problem}

\newcommand{\shepauthor}
{S. F. King\footnote{talk given by S.F.King}  and I. N. R. Peddie}

\newcommand{\shepaddress}{School of Physics and Astronomy,\\ 
  University of Southampton, Southampton, SO17 1BJ, U.K.\\ 
  E-mail: king@soton.ac.uk, inrp@soton.ac.uk}

\newcommand{\shepabstract}
{
  We consider the problem of trying to understand the recently measured
  neutrino data simultaneously with understanding the heirarchical form
  of quark and charged lepton Yukawa matrices. We summarise the data that
  a sucessful model of neutrino mass must predict, and then move on to 
  attempting to do so in the context of spontaneously broken `family' 
  symmetries. We consider first an abelian $U(1)$ family symmetry, which appears
  in the context of a type I string model. Then we consider a model based
  on a non-abelian $SU(3)_F$, which is the maximal family group consitent
  with an $SO(10)$ GUT. In this case the symmetry is more constraining, and
  is examined in the context of SUSY field theory.
}

\begin{document}

\begin{titlepage}
  \begin{flushright}   
    hep-ph/0312235 \\
    SHEP/0335
  \end{flushright}
  \begin{center}
    {\large{\bf \sheptitle} } \\
    \shepauthor \\
    \mbox{} \\
    {\it \shepaddress }  \\
    \vskip 0.1in
    {\bf Abstract } \\
     \bigskip    
  \end{center}
\setcounter{page}{0}
\shepabstract
\begin{flushleft}
  \today
\end{flushleft}
\end{titlepage}

\section{Introduction}

Since the publication of results by SNO \cite{SNO:2002}
and KamLAND  \cite{KamLAND:2002}, we now have a reasonably
good picture of the neutrino sector of the Standard Model.
In fact, by far the best fit to the data is the Large Mixing Angle (LMA) MSW
fit \cite{deHolanda:2003nj}
We only have measurements of the mass differences in the neutrino
sector:
\begin{equation}
  \label{eq:1}
  \Delta m^2_{21} = (0.008 \mathrm{eV})^2\;\;\; 
  \left| \Delta m^2_{32} \right| = ( 0.05 \mathrm{eV} )^2
\end{equation}

There are three possible neutrino mass patterns consistent with LMA \cite{King:2003jb}.
The first, ``normal'', has $\Delta m^2_{32} > 0$ and $m_1 \approx 0$.
The second, ``inverted'', has $\Delta m^2_{32} < 0$ and $m_3 \approx 0$.
The third, ``quasi-degenerate'', has the neutrino masses at a scale
where the mass differences are negligable $m_1 \approx m_2 \approx m_3$.

We also have two measurably large neutrino mixing angles, and one small mixing
angle:
\begin{equation}
  \label{eq:13}
  \theta_{\mathrm{sol}} \equiv \theta_{12} \approx \frac{\pi}{6} \;\;\;
  \theta_{\mathrm{atm}} \equiv \theta_{23} \approx \frac{\pi}{4} \;\;\;
  \theta_{\mathrm{CHOOZ}} \equiv \theta_{13} \le 0.2
\end{equation}

In the Standard Model, neutrinos are massless, and neutrinos and anti-neutrinos
are distinguished by a total conserved lepton number, $L$. Since we now know
that this is not true, we wish to understand why the neutrino masses are so much
smaller than that of the quarks and the charged leptons. Related to this is whether
they have a Majorana or Dirac mass term, or both.

\section{Seesaw models}
\label{sec:seesaw-models}

Dirac mass terms are just like mass terms for the charged leptons and conserve
lepton number. They can follow from a neutrino Yukawa coupling, just like charged
lepton masses do in the Standard Model:
\begin{equation}
  \label{eq:29}
  m_{LR} \overline{\nu_L} \nu_R
\end{equation}
Majorana mass terms violate lepton number:
\begin{equation}
  \label{eq:30}
  m_{LL} \overline{\nu_L} \nu_L^{(c)}
\end{equation}
\begin{equation}
  M_{RR} \overline{\nu_R} \nu_R^{(c)}
  \label{eq:31}  
\end{equation}

The term $m_{LL}$ violates the electroweak gauge symmetry, so we would expect
it to be exactly zero. However, no symmetry exists which protects $m_{RR}$, so
we would expect that to be very heavy, of the order of $10^{16}\;\;\mathrm{GeV}$.
If we have both Dirac and Majorana mass terms, then we can generate small masses
of the order of $\Delta m^2$, by a Type I seesaw mechanism.

\begin{equation}
  \label{eq:32}
  \left( 
    \begin{array}{cc}
      \overline{\nu_L} & \overline{\nu_R^{(c)}}
    \end{array}
  \right)
  \left(
    \begin{array}{cc}
      0 & m_{LR} \\
      m_{LR}^T & M_{RR}
    \end{array}
  \right)
  \left(
    \begin{array}{c}
      \nu_L^{(c)} \\
      \nu_R
    \end{array}
  \right)
\end{equation}

In order to get the physical masses, we must block diagonalise this matrix. Assuming
that $M_{RR} \gg m_{LR}$, we find
\begin{equation}
  \label{eq:33}
  m_{LL}^\prime \approx m_{LR} M_{RR}^{-1} m_{LR}^T
\end{equation}
\begin{equation}
  \label{eq:34}
  M_{RR}^\prime \approx M_{RR}
\end{equation}

We can then enumerate the forms of $m_{LL}$ that are consitent with LMA MSW. 
We refer to terms with a zero in the 11 element `type A', and those without a
zero in the 11 element `type B'. 
There is one possiblity with a ``normal'' heirarchy ($m_1^2 \ll m_2^2 \ll m_3^2$):
\begin{equation}
  \label{eq:35}
  m_{LL}^{HI,A} \approx
  \left(
    \begin{array}{ccc}
      0 & 0 & 0 \\
      0 & 1 & 1 \\
      0 & 1 & 1 
    \end{array}
  \right) \frac{m}{2}
\end{equation}
There are two possibilities with an ``inverted'' heirarchy ($m_1^2 \sim m_2^2 \gg m_3^2$):
\begin{equation}
  \label{eq:36}
  m_{LL}^{IH,A} \approx
  \left(
    \begin{array}{ccc}
    0 & 1 & 1 \\
    1 & 0 & 0 \\
    1 & 0 & 0
  \end{array}
\right)
  \frac{m}{\sqrt{2}}\;\;\;\;\;
  m_{LL}^{IH,B} \approx
  \left(
    \begin{array}{ccc}
      1 & 0 & 0 \\
      0 & \frac{1}{2} & \frac{1}{2} \\
      0 & \frac{1}{2} & \frac{1}{2} 
    \end{array}
  \right) m
\end{equation}
There are three possibilities with a degenerate mass pattern ($m_1^2 \approx m_2^2 \approx m_3^2$):
\begin{displaymath}
  m_{LL}^{DEG,A} \approx
  \left(
    \begin{array}{ccc}
      0 & \frac{1}{\sqrt{2}} & \frac{1}{\sqrt{2}} \\
      \frac{1}{\sqrt{2}} & \frac{1}{2} & \frac{1}{2} \\
      \frac{1}{\sqrt{2}} & \frac{1}{2} & \frac{1}{2}
    \end{array}
  \right)m\;\;\;\;\;
  m_{LL}^{DEG,B1} \approx
  \left(
    \begin{array}{ccc}
      1 & 0 & 0 \\
      0 & 1 & 0 \\
      0 & 0 & 1
    \end{array}
  \right)m
\end{displaymath}
\begin{equation}
  \label{eq:37}
  m_{LL}^{DEG,B2} \approx
  \left(
    \begin{array}{ccc}
      1 & 0 & 0 \\
      0 & 0 & 1 \\
      0 & 1 & 0
    \end{array}
  \right) m
\end{equation}

$m_{LL}^{IH,B}$ leads to a large rate for neutrinoless double beta decay.

From this point on, we focus on the normal heirarchy. In this case, we need
to understand why $m_2^2 \ll m_3^2$, $\theta_{23} \approx \frac{\pi}{4}$,
$\theta_{12} \approx \frac{\pi}{6}$. The technical requirement for $m_2^2 \ll m_3^2$
is for the subdeterminant to be small:
\begin{equation}
  \label{eq:38}
  \mathrm{det}
  \left|
    \begin{array}{cc}
      m_{22} & m_{23} \\
      m_{32} & m_{33} 
    \end{array}
  \right|
  \ll m^2
\end{equation}

We are then led to ask why this sub-determinant is small, and why the solar
angle is large. One model which solves this is right-handed neutrino dominance.

If one right-handed neutrino dominates in the see-saw mechanism, and couples
equally to the second and third family left-handed neutrinos then $m_2^2 \ll m_3^2$
and $\theta_{23} \approx \frac{\pi}{4}$ \cite{King:SRHND}

Furthermore, if a second right-handed neutrino gives the leading sub-dominant
contribuitions to the see-saw mechanims and couples equally to all three 
left-handed generations, then a large solar angle is gnerated 
$\theta_{12} \approx \frac{\pi}{6}$\cite{King:SeqDom}

The corollary of this is that if the dominant right-handed neutrino is the
lightest, then there is a link between the neturino oscillation phase and
the phases of leptogenesis and neutrinoless double beta decay \cite{King:2002qh}.

\section{The flavour problem}
\label{sec:flavour-problem}

There are two parts to the flavour problem. This fist is understanding
the origin of the Yukawa couplings, (and heavy Majorana masses for the
see-saw mechanism), which lead to low energy quark and lepton mixing angles.
In low energy SUSY, we also need to understand why flavour changing and/or
CP violating processes induced by SUSY loops are so small. A theory of
flavour must address both these problems simultaneously.

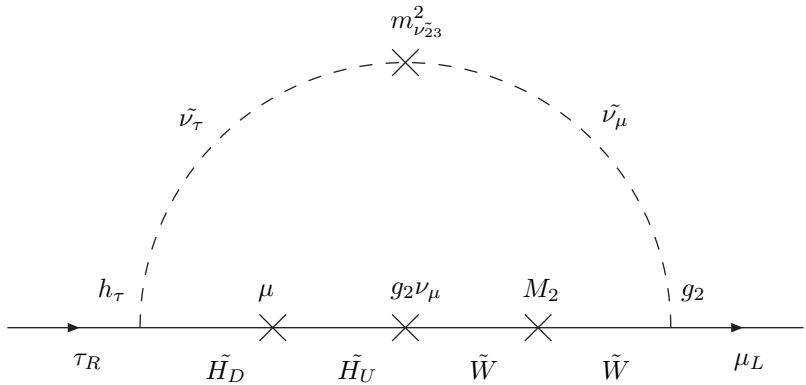
\begin{figure}[htbp]
  \centering
  \begin{picture}(350,140)
    \ArrowLine(25,20)(75,20)
    \Text(50,10)[tl]{$\tau_R$}
    \Line(75,20)(125,20)
    \Text(100,10)[tl]{$\tilde{H_D}$}
    \Line(125,20)(175,20)
    \Text(150,10)[tl]{$\tilde{H_U}$}
    \Line(175,20)(225,20)
    \Text(200,10)[tl]{$\tilde{W}$}
    \Line(225,20)(275,20)
    \Text(250,10)[tl]{$\tilde{W}$}
    \ArrowLine(275,20)(325,20)
    \Text(300,10)[tl]{$\mu_L$}
    \Line(120,15)(130,25)
    \Line(120,25)(130,15)
    \Text(120,30)[bl]{$\mu$}
    \Line(170,15)(180,25)
    \Line(170,25)(180,15)
    \Text(170,30)[bl]{$g_2 \nu_\mu$}
    \Line(220,15)(230,25)
    \Line(220,25)(230,15)
    \Text(220,30)[bl]{$M_2$}
    \Text(70,30)[br]{$h_\tau$}
    \Text(280,30)[bl]{$g_2$}
    \DashCArc(175,20)(100,0,180){5}
    \Line(170,115)(180,125)
    \Line(170,125)(180,115)
    \Text(170,130)[bl]{$m^2_{\tilde{\nu_{23}}}$}
    \Text(100,95)[br]{$\tilde{\nu_\tau}$}
    \Text(250,95)[bl]{$\tilde{\nu_\mu}$}
  \end{picture}
  
  \caption{Example of a SUSY loop contributing to $\tau\rightarrow\mu\gamma$}
  \label{fig:susy_loop_tmg}
\end{figure}

Consider, for example the loop in fig.~\ref{fig:susy_loop_tmg}. This leads
\cite{taumugamma}
to a rate:
\begin{equation}
  \label{eq:39}
  BR(\tau\rightarrow\mu\gamma) \approx \frac{\alpha^3}{G_F^2} f_{32}(M_2,\mu,m_{\tilde{\nu}})
  \left| m^2_{\tilde{L}_{32}}\right|^2 \tan^2\beta
\end{equation}

We see that the decay rate depends on off-diagonal slepton masses. There
will be two sources of slepton masses. The first is `primordial'; this is
where there are off-diagonal elements in the SCKM basis at the high-energy
scale, generated by the SUSY breaking mechanism.
The second is RGE generated. This can be from running a GUT theory from the
Planck scale to the GUT scale if Higgs triplets are present. It can also come
from running the MSSM with right-handed neutrinos from the Planck scale to the
lightest right-handed neutrino mass scale.

In general both sources will be present.

We can address both problems at the same time by employing a family symmetry
\cite{Froggatt:1978nt}. This will be spontanesouly by $\Phi$, a Higgs field for the
family symmetry. The idea is that each generation of matter will not be neutral
under the symmetry, and so extra powers of the flavon will appear over some
UV cutoff. This will lead to effective Yukawa couplings when $\Phi$ gains
a VEV:

\begin{equation}
  \label{eq:40}
  \psi^i \overline{\psi^j} H \left(\frac{\Phi}{M}\right)^{n_{ij}}
  \rightarrow \psi^i \overline{\psi^j} H \left( \frac{\left<\Phi\right>}{M}\right)^{n_{ij}}
  \;\;\;\;\;
  \frac{\left<\Phi\right>}{M} \sim 0.1
\end{equation}

This gives an explanation of the Yukawa textures:

\begin{equation}
  \label{eq:41}
  Y = 
  \left(
    \begin{array}{ccc}
      \left(\frac{\Phi}{M}\right)^{n_{11}} &
      \left(\frac{\Phi}{M}\right)^{n_{12}} &
      \left(\frac{\Phi}{M}\right)^{n_{13}} \\
      \left(\frac{\Phi}{M}\right)^{n_{21}} &
      \left(\frac{\Phi}{M}\right)^{n_{22}} &
      \left(\frac{\Phi}{M}\right)^{n_{23}} \\
      \left(\frac{\Phi}{M}\right)^{n_{31}} &
      \left(\frac{\Phi}{M}\right)^{n_{32}} &
      \left(\frac{\Phi}{M}\right)^{n_{33}}
    \end{array}
  \right)
\end{equation}

This addresses the first part of the flavour problem. However,
it can also make the second part problematic. This is because
the new fields in the Yukawa operators can develop F-term VEVs,
and contribute to the SUSY breaking F-terms in a non-universal
way. This leads to a new and dangerous source of primordial
flavour violation \cite{Atermstuff,King:2003kf}:
\begin{equation}
  \label{eq:42}
  \Delta A = F_\Phi \partial_\Phi \ln \Phi^n = F_\phi \frac{n}{\Phi}
\end{equation}
But the auxilliary field is proportional to the scalar component:
\begin{equation}
  \label{eq:43}
  F_\Phi \propto m_{3/2} \Phi \rightarrow \Delta A \propto n m_{3/2}
\end{equation}

And example of this with a $U(1)$ family symmetry is:
\begin{equation}
  \label{eq:41}
  Y = 
  \left(
    \begin{array}{ccc}
      \left(\frac{\Phi}{M}\right)^5 &
      \left(\frac{\Phi}{M}\right)^3 &
      \left(\frac{\Phi}{M}\right) \\
      \left(\frac{\Phi}{M}\right)^4 &
      \left(\frac{\Phi}{M}\right)^2 &
      1 \\
      \left(\frac{\Phi}{M}\right)^4 &
      \left(\frac{\Phi}{M}\right)^2 &
      1
    \end{array}
  \right)
  \rightarrow \Delta A \sim m_{3/2} 
  \left(
    \begin{array}{ccc}
      5 & 3 & 1 \\
      4 & 2 & 0 \\
      4 & 2 & 0
    \end{array}
  \right)
\end{equation}

If we take a specific model where we can switch off
the new effects, and look at more standard SUGRA flavour
violation, we can gauge the relative importance of the
new effects \cite{King:2003kf}. In order to do so, we
look at three benchmark points. Point A corresponds to
minimum flavour violation, where the SUGRA setup is like
mSUGRA. In this case the seesaw RGE contributions are
the only contributions. Point B corresponds to a `standard'
SUGRA FV setup, where non-universal scalar masses generate
primordial FV in the SCKM basis. Finally, point C corresponds
to the new effects, where $\Delta A$ generates primordial FV
even before switching to the SCKM basis.

\begin{figure}[htbp]
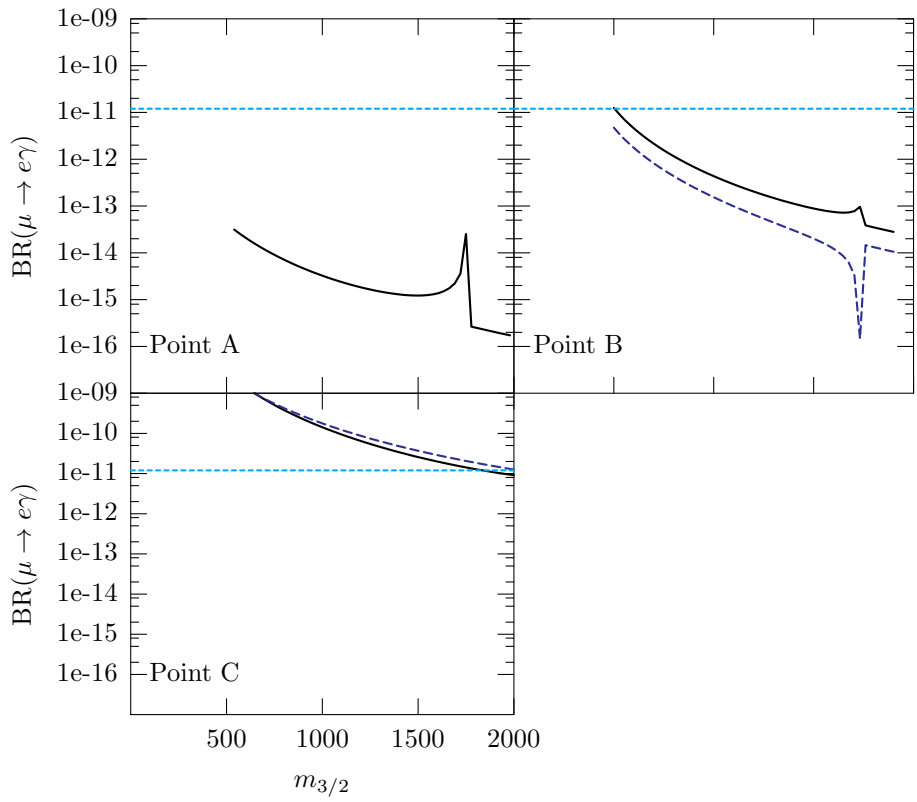

  \centering
  \include{m32_meg_lmeg}
  \caption{$BR(\mu\rightarrow e\gamma)$ for the three benchmark points}
  \label{fig:graphs_meg}
\end{figure}

\begin{figure}[htbp]
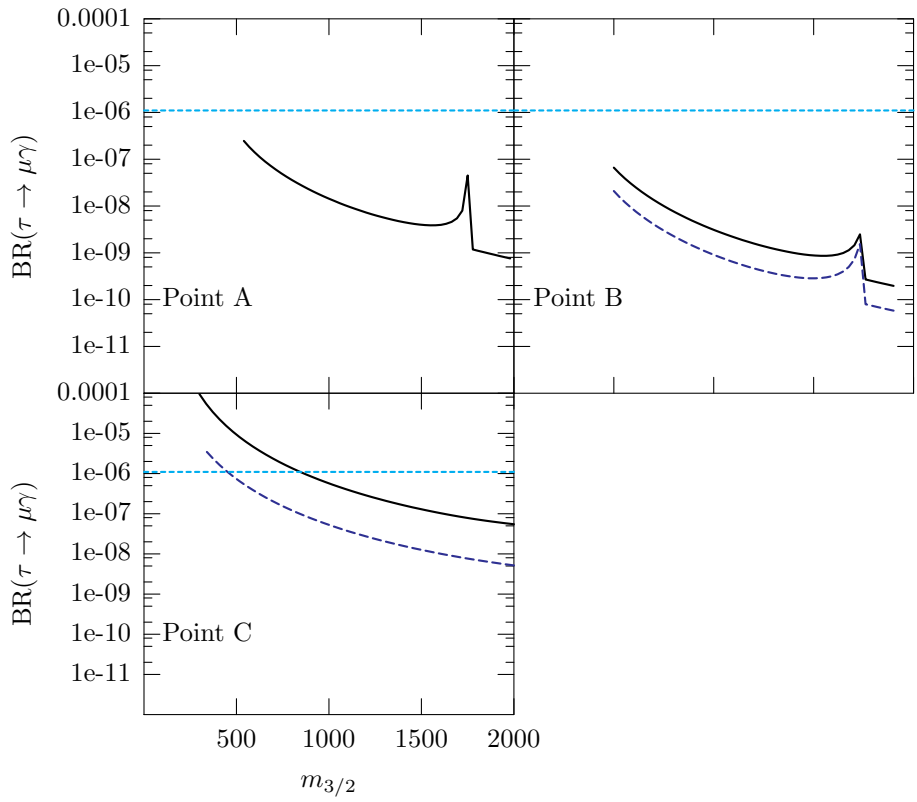

  \centering
  \include{m32_tmg}
  \caption{$BR(\tau\rightarrow\mu\gamma)$ for the three benchmark points}
  \label{fig:graphs_tmg}
\end{figure}

We display, in fig.~\ref{fig:graphs_meg} $BR(\mu\rightarrow e\gamma)$
at the three sesesaw points, and in fig.~\ref{fig:graphs_tmg} 
$BR(\tau\rightarrow\mu\gamma)$ at the three benchmark points.

There is no reason why a family symmetry has to be abelian.
Consider a SUSY family GUT $SO(10)_G \otimes SU(3)_F$ \cite{King:2003rf}. 
$SU(3)$ is the largest family symmetry consitent with a $SO(10)$ GUT.
This model is an example of sequential dominance, gives an excellent
description of the quark and lepton masses and mixing angles, and
can address the SUSY flavour/CP problems.

In this model, we break the $SO(10)$ GUT in the Pati-Salam direction
by a Wilson line breaking. The Pati-Salm group is also broken to the MSSM
group by Wilson line breaking. The $SU(3)_F$ is broken first to $SU(2)_F$
by a Higgs field $\phi_3$. The remnant $SU(2)_F$ is then broken completely
by another Higgs field $\phi_{23}$.

There are a few global symmetries in the theory to restrict the Yukawa
operators allowed. The leading order operator leads to the top Yukawa
element ( $\overline\epsilon \approx 0.15\,,\, \epsilon \approx 0.05$):
\begin{equation}
  \label{eq:44}
  \frac{1}{M^2} \psi \phi_3 \psi \phi_3 h 
  \rightarrow
  Y = 
  \left(
    \begin{array}{ccc}
      0 \\
      & 0 \\
      & & 1
    \end{array}
  \right) \overline\epsilon
\end{equation}
The subleading operator leads to the charm Yukawa and the charm-top mixing
angles:
\begin{equation}
  \label{eq:45}
  \frac{\Sigma}{M^3} \psi \phi_{23} \psi \phi_{23} h
  \rightarrow Y = 
  \left(
    \begin{array}{ccc}
      0  \\
      & y \epsilon^2 & y \epsilon^2 \\
      & y \epsilon^2 & y \epsilon^2 
    \end{array}
  \right)
\end{equation}

Setting the $\mathcal{O}(1)$ coefficients of the operators to get a good fit,
we predict the following Yukawa matrices:

\begin{equation}
  \label{eq:24}
  Y^u \sim 
  \left(
    \begin{array}{ccc}
      0 & 1.2\epsilon^3 & 0.9\epsilon^3 \\
      -1.2\epsilon^3 & -\frac{2}{3}\epsilon^2 & -\frac{2}{3}\epsilon^2 \\
      -0.9\epsilon^3 &-\frac{2}{3}\epsilon^2 & 1
    \end{array}
  \right)\overline{\epsilon}\;,\;
  Y^d \sim
  \left(
    \begin{array}{ccc}
      0 & 1.6\overline{\epsilon}^3 & 0.7 \overline{\epsilon}^3 \\
      -1.6\overline{\epsilon^3} & \overline{\epsilon}^2 
      & \overline{\epsilon}^3 + \overline{\epsilon}^\frac{5}{2} \\
      -0.7\overline{\epsilon}^3 
      & \overline{\epsilon}^2 - \overline{\epsilon}^\frac{5}{2} & 1     
    \end{array}
  \right)\overline{\epsilon}
\end{equation}
\begin{equation}
  \label{eq:25}
  Y^\nu \sim 
  \left(
    \begin{array}{ccc}
      0 & 1.2\epsilon^2 & 0.9\epsilon^2 \\
      -1.2\epsilon^2 & -\alpha\epsilon^2 
      & -\alpha\epsilon^2 
      +\frac{\epsilon^3}{\sqrt{\overline{\epsilon}}} \\
      -0.9\epsilon^3 & -\alpha\epsilon^2 
      - \frac{\epsilon^3}{\sqrt{\overline{\epsilon}}} & 1
    \end{array}
  \right)\overline{\epsilon}
  \;,\;
  Y^e \sim
  \left(
    \begin{array}{ccc}
    0 & 1.6\overline{\epsilon}^3 & 0.7\overline{\epsilon}^3 \\
    -1.6\overline{\epsilon}^3 & 3\overline{\epsilon}^2 
    & 3\overline{\epsilon}^2 \\
    -0.7\overline{\epsilon}^3 & 3\overline{\epsilon}^2 & 1
\end{array}
  \right)\overline{\epsilon}
\end{equation}
\begin{equation}
  \label{eq:26}
  \frac{M_{RR}}{M_3} \sim
  \left(
    \begin{array}{ccc}
      \epsilon^6\overline{\epsilon}^3 & & \\
      & \epsilon^6\overline{\epsilon}^2 & \\
      & & 1
    \end{array}
  \right)
\end{equation}

The first RH neutrino dominates, and we predict 
$m_2/m_3 \sim \overline{\epsilon}$, $\tan\theta_{23} \sim 1.3$, 
$\tan\theta_{12} \sim 0.66$ and $\theta_{13} \sim \overline{\epsilon}$.

This all assumes a canonical K\"ahler metric, and so we would
expect the soft SUSY breaking masses to be universal for a simple
SUSY breaking scenario.

\section{Conclusions}
\label{sec:conclusions}

Small neutrino masses can be elegantly explained by the see-saw mechanism.
In that case, sequential dominance then provides a natural explanation of a neutrino
mass heirarchy and large mixing angles. If the dominant right-handed
neutrino is the lightest one, there is a link between the leptogensis phase
and the CP phase which is measurable at a neutrino factory.

Family symmetries provide a natural way of understanding the heirarchies
in the quark and charged lepton masses, and the smallness of the quark
mixing angles. In this case, dangerous new sources of flavour changing
masses in general arise from Yukawa operators which lead to large off-diagonal
soft trilinears. One example is the $U(1)$ family symmetry.
Another is the $SU(3)$ family symmetry, which provides an excellent description
of the fermion spectrum, with SUSY flavour-changing controlled by the family
symmetries.

\end{document}

%% file: m32_meg_lmeg.tex
\begin{picture}(0,0)%
\includegraphics{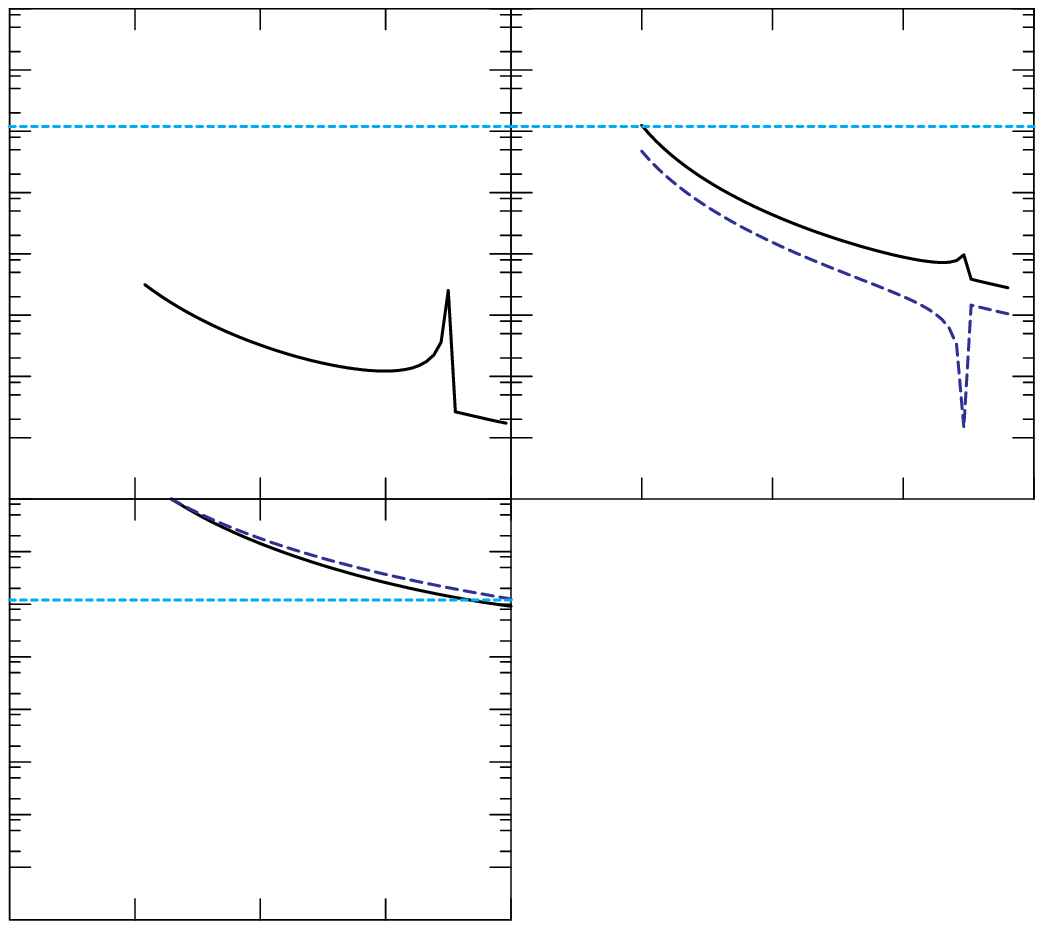}%
\end{picture}%
\setlength{\unitlength}{0.0200bp}%
\begin{picture}(18000,15119)(0,0)%
\put(2250,2257){\makebox(0,0)[r]{\strut{}1e-16}}%
\put(2250,3015){\makebox(0,0)[r]{\strut{}1e-15}}%
\put(2250,3772){\makebox(0,0)[r]{\strut{}1e-14}}%
\put(2250,4530){\makebox(0,0)[r]{\strut{}1e-13}}%
\put(2250,5287){\makebox(0,0)[r]{\strut{}1e-12}}%
\put(2250,6045){\makebox(0,0)[r]{\strut{}1e-11}}%
\put(2250,6802){\makebox(0,0)[r]{\strut{}1e-10}}%
\put(2250,7560){\makebox(0,0)[r]{\strut{}1e-09}}%
\put(4305,1000){\makebox(0,0){\strut{}500}}%
\put(6110,1000){\makebox(0,0){\strut{}1000}}%
\put(7915,1000){\makebox(0,0){\strut{}1500}}%
\put(9720,1000){\makebox(0,0){\strut{}2000}}%
\put(500,4530){\rotatebox{90}{\makebox(0,0){\strut{}$\mathrm{BR}(\mu\rightarrow e\gamma)$}}}%
\put(6110,250){\makebox(0,0){\strut{}$m_{3/2}$}}%
\put(2861,2257){\makebox(0,0)[l]{\strut{}Point C}}%
\put(2250,8442){\makebox(0,0)[r]{\strut{}1e-16}}%
\put(2250,9325){\makebox(0,0)[r]{\strut{}1e-15}}%
\put(2250,10207){\makebox(0,0)[r]{\strut{}1e-14}}%
\put(2250,11090){\makebox(0,0)[r]{\strut{}1e-13}}%
\put(2250,11972){\makebox(0,0)[r]{\strut{}1e-12}}%
\put(2250,12855){\makebox(0,0)[r]{\strut{}1e-11}}%
\put(2250,13737){\makebox(0,0)[r]{\strut{}1e-10}}%
\put(2250,14620){\makebox(0,0)[r]{\strut{}1e-09}}%
\put(4305,7060){\makebox(0,0){\strut{}}}%
\put(6110,7060){\makebox(0,0){\strut{}}}%
\put(7915,7060){\makebox(0,0){\strut{}}}%
\put(9720,7060){\makebox(0,0){\strut{}}}%
\put(500,11090){\rotatebox{90}{\makebox(0,0){\strut{}$\mathrm{BR}(\mu\rightarrow e\gamma)$}}}%
\put(2861,8442){\makebox(0,0)[l]{\strut{}Point A}}%
\put(9470,8442){\makebox(0,0)[r]{\strut{}}}%
\put(9470,9325){\makebox(0,0)[r]{\strut{}}}%
\put(9470,10207){\makebox(0,0)[r]{\strut{}}}%
\put(9470,11090){\makebox(0,0)[r]{\strut{}}}%
\put(9470,11972){\makebox(0,0)[r]{\strut{}}}%
\put(9470,12855){\makebox(0,0)[r]{\strut{}}}%
\put(9470,13737){\makebox(0,0)[r]{\strut{}}}%
\put(9470,14620){\makebox(0,0)[r]{\strut{}}}%
\put(11603,7060){\makebox(0,0){\strut{}}}%
\put(13485,7060){\makebox(0,0){\strut{}}}%
\put(15368,7060){\makebox(0,0){\strut{}}}%
\put(17250,7060){\makebox(0,0){\strut{}}}%
\put(10097,8442){\makebox(0,0)[l]{\strut{}Point B}}%
\end{picture}%
 

%% file: m32_tmg.tex
\begin{picture}(0,0)%
\includegraphics{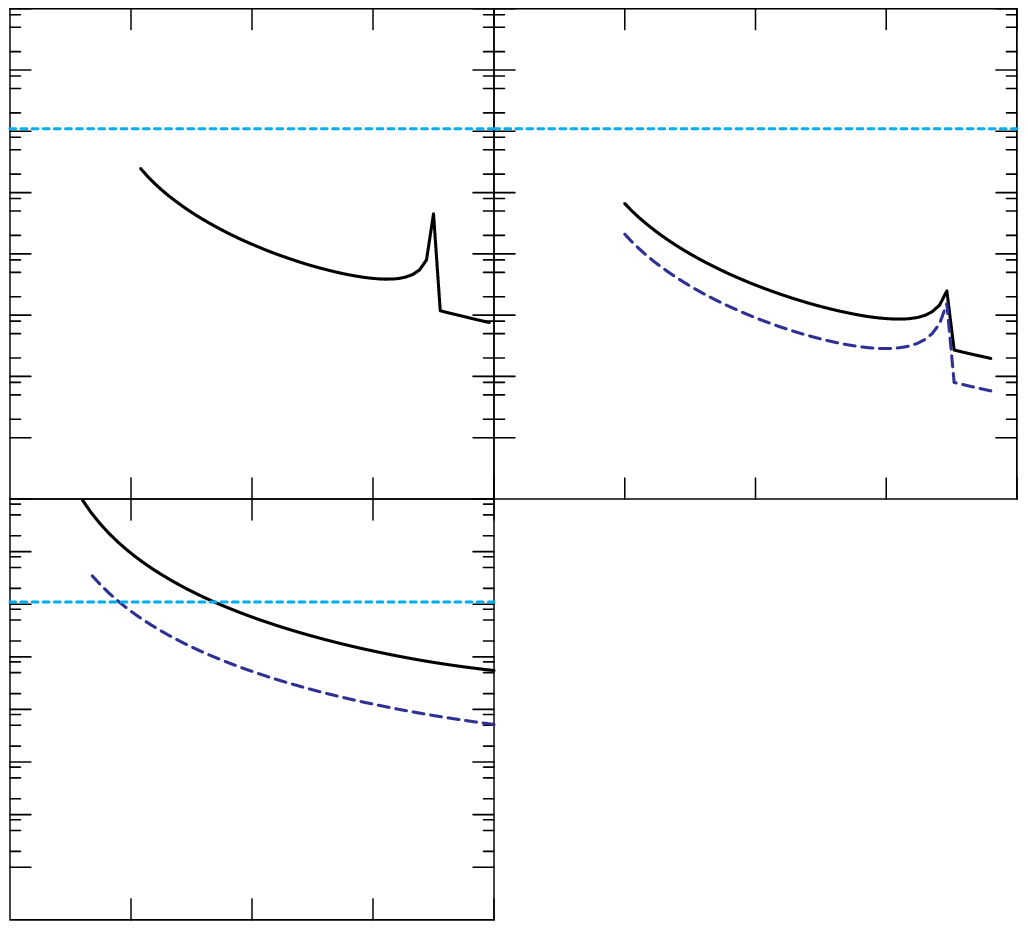}%
\end{picture}%
\setlength{\unitlength}{0.0200bp}%
\begin{picture}(18000,15119)(0,0)%
\put(2500,2257){\makebox(0,0)[r]{\strut{}1e-11}}%
\put(2500,3015){\makebox(0,0)[r]{\strut{}1e-10}}%
\put(2500,3772){\makebox(0,0)[r]{\strut{}1e-09}}%
\put(2500,4530){\makebox(0,0)[r]{\strut{}1e-08}}%
\put(2500,5287){\makebox(0,0)[r]{\strut{}1e-07}}%
\put(2500,6045){\makebox(0,0)[r]{\strut{}1e-06}}%
\put(2500,6802){\makebox(0,0)[r]{\strut{}1e-05}}%
\put(2500,7560){\makebox(0,0)[r]{\strut{}0.0001}}%
\put(4492,1000){\makebox(0,0){\strut{}500}}%
\put(6235,1000){\makebox(0,0){\strut{}1000}}%
\put(7977,1000){\makebox(0,0){\strut{}1500}}%
\put(9720,1000){\makebox(0,0){\strut{}2000}}%
\put(500,4530){\rotatebox{90}{\makebox(0,0){\strut{}$\mathrm{BR}(\tau\rightarrow\mu\gamma)$}}}%
\put(6235,250){\makebox(0,0){\strut{}$m_{3/2}$}}%
\put(3098,3015){\makebox(0,0)[l]{\strut{}Point C}}%
\put(2500,8442){\makebox(0,0)[r]{\strut{}1e-11}}%
\put(2500,9325){\makebox(0,0)[r]{\strut{}1e-10}}%
\put(2500,10207){\makebox(0,0)[r]{\strut{}1e-09}}%
\put(2500,11090){\makebox(0,0)[r]{\strut{}1e-08}}%
\put(2500,11972){\makebox(0,0)[r]{\strut{}1e-07}}%
\put(2500,12855){\makebox(0,0)[r]{\strut{}1e-06}}%
\put(2500,13737){\makebox(0,0)[r]{\strut{}1e-05}}%
\put(2500,14620){\makebox(0,0)[r]{\strut{}0.0001}}%
\put(4492,7060){\makebox(0,0){\strut{}}}%
\put(6235,7060){\makebox(0,0){\strut{}}}%
\put(7977,7060){\makebox(0,0){\strut{}}}%
\put(9720,7060){\makebox(0,0){\strut{}}}%
\put(500,11090){\rotatebox{90}{\makebox(0,0){\strut{}$\mathrm{BR}(\tau\rightarrow\mu\gamma)$}}}%
\put(3098,9325){\makebox(0,0)[l]{\strut{}Point A}}%
\put(9470,8442){\makebox(0,0)[r]{\strut{}}}%
\put(9470,9325){\makebox(0,0)[r]{\strut{}}}%
\put(9470,10207){\makebox(0,0)[r]{\strut{}}}%
\put(9470,11090){\makebox(0,0)[r]{\strut{}}}%
\put(9470,11972){\makebox(0,0)[r]{\strut{}}}%
\put(9470,12855){\makebox(0,0)[r]{\strut{}}}%
\put(9470,13737){\makebox(0,0)[r]{\strut{}}}%
\put(9470,14620){\makebox(0,0)[r]{\strut{}}}%
\put(11603,7060){\makebox(0,0){\strut{}}}%
\put(13485,7060){\makebox(0,0){\strut{}}}%
\put(15368,7060){\makebox(0,0){\strut{}}}%
\put(17250,7060){\makebox(0,0){\strut{}}}%
\put(10097,9325){\makebox(0,0)[l]{\strut{}Point B}}%
\end{picture}%
 